\documentclass[twocolumn,aps,showpacs]{revtex4}
\usepackage{amsfonts}
\usepackage{amsmath}
\usepackage{graphicx}
\usepackage{bm}
\usepackage{amssymb}
\usepackage{dcolumn}

\newcommand{\bk}{\mbox{\boldmath $k$}}

\begin{document}

\title{
Combined approach of density functional theory and quantum Monte Carlo method\\
to electron correlation in dilute magnetic semiconductors}

\author{Jun-ichiro Ohe$^1$, Yoshihiro Tomoda$^1$, Nejat Bulut$^{1,2}$,
Ryotaro Arita$^3$, Kazuma Nakamura$^3$,\\
and Sadamichi Maekawa$^{1,2}$}

\affiliation{
$^1$Institute for Materials Research, Tohoku University, Sendai 980-8577, Japan\\
$^2$CREST, Japan Science and Technology Agency (JST), Kawaguchi, Saitama 332-0012, Japan\\
%$^3$Condensed Matter Theory Laboratory, RIKEN, Wako, Saitama 351-0198, Japan
$^3$Department of Applied Physics, University of Tokyo, Bunkyo-ku,
Tokyo 113-8656, Japan
}

\date{\today}

\begin{abstract}

We present a realistic study for electronic and magnetic properties 
in dilute magnetic semiconductor (Ga,Mn)As.
A multi-orbital Haldane-Anderson model parameterized by density-functional 
calculations is presented and solved with the Hirsch-Fye 
quantum Monte Carlo algorithm. 
Results well reproduce experimental results in the dilute limit. 
When the chemical potential is located between the top of the valence band and 
an impurity bound state, a long-range 
ferromagnetic correlations between the impurities, 
mediated by antiferromagnetic impurity-host couplings, 
are drastically developed.
We observe an anisotropic character in local density of states 
at the impurity-bound-state energy, which is consistent with the STM measurements.
The presented combined approach thus offers a firm starting point 
for realistic calculations of the various family of dilute magnetic
semiconductors.  

\end{abstract}

\pacs{75.50.Pp, 75.30.Hx, 75.40.Mg, 71.55.-i}

\maketitle

The discovery of ferromagnetism in dilute magnetic semiconductor
(DMS) materials, represented by a (Ga,Mn)As system, has created 
much activity in the field of spintronics 
\cite{Maekawa2006, Ohno1992, Zutic2004, Jungwirth2006}. 
In the low-doping regime ($\ll 1\%$), (Ga,Mn)As is
insulating with a clear experimental evidence for the presence of an
Mn-induced impurity band located at 110 meV above the valence band
\cite{Jungwirth2007}. 
The position of the impurity level and 
the impurity-induced carriers 
 are key quantities in generating 
the ferromagnetism.  
Various model studies  
\cite{Ichimura2006,Krstajic,Yamauchi,Inoue,Bulut2007} 
have clarified that this ferromagnetic correlations 
between the impurities are based on the polarization mechanism of the
impurity-induced carriers.

 The challenge for predicting new functional properties of the DMS materials
 and their realistic designs clearly requires many tasks such as 
 nonempirical electronic-band structure of the host compound,  
 reliable estimations for impurity-host hybridizations,  
 and rigorous treatments for local electron correlations 
 at the impurity site. 
 A combined approach of 
density functional theory (DFT) \cite{Ref_DFT} and 
 quantum Monte Carlo (QMC) technique \cite{Hirsch1986} 
 is clearly useful for this purpose. 
The hybridization or mixing parameters between the host and 
impurity orbitals is a crucial parameter characterizing 
 material differences and therefore this parameter should be estimated 
on the basis of {\em ab initio} density-functional calculations. 
On the other hand, density-functional calculations often fail to describe 
local electron correlations, so this problem 
 must be treated with more accurate solvers such as the QMC technique. 
Critical comparisons with the experimental data 
are needed for a check 
whether this approach is really reliable in describing 
the details of the "low-energy" electronic and magnetic properties 
and/or whether it has an extended applicability for many DMS materials.

\begin{figure}[bp]
\begin{center}
\includegraphics[scale=0.6]{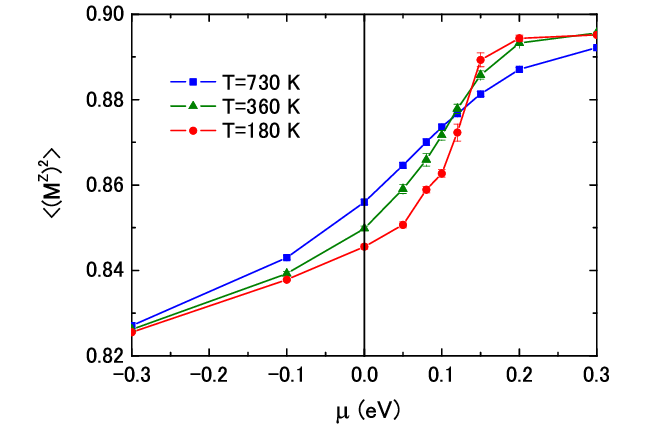}
\caption{(Color online) Square of the impurity magnetic moment
$\langle (M^z)^2\rangle$ {\em vs} the chemical potential $\mu$ at
various temperatures in the single-impurity case. These results
are for the single-impurity Haldane-Anderson model.
The vertical line ($\mu=0$) denotes the top of the valence band.
\label{fig1}}
\end{center}
\end{figure}

In this letter, we present a comprehensive study 
for the generation mechanism of the ferromagnetism 
of a dilute (Ga,Mn)As system. 
A multi-orbital Haldane-Anderson model \cite{Haldane1976}
is employed to describe electronic/magnetic properties 
of this system.
First principles band-structure calculations are performed to 
evaluate the mixing matrix elements in this model and   
the resulting model is solved with the help of the QMC technique 
to examine the inter-impurity and impurity-host magnetic correlations. 
The hybridization between host and impurity electrons generates the impurity bound state \cite{Blakemore1973}.
When the chemical potential is located between 
the impurity bound state and the valence-band top, 
a notable long-range ferromagnetic correlation between the impurities 
and its dramatic enhancement as the temperature decreases are observed. 
We will show that this ferromagnetic correlation is mediated by 
antiferromagnetic impurity-host couplings. 
In addition, our calculated host local density of states 
is highly anisotropic around the impurity sites, 
which is also consistent with the experimental STM image \cite{Yakunin2004}. 
These quantitative agreements with the experiments 
indicates that the presented combination idea is a useful approach   
for aiming at realistic calculations and 
searching new functional high-$T_c$ DMS materials.

\begin{figure}[tbp]
\begin{center}
\includegraphics[scale=0.6]{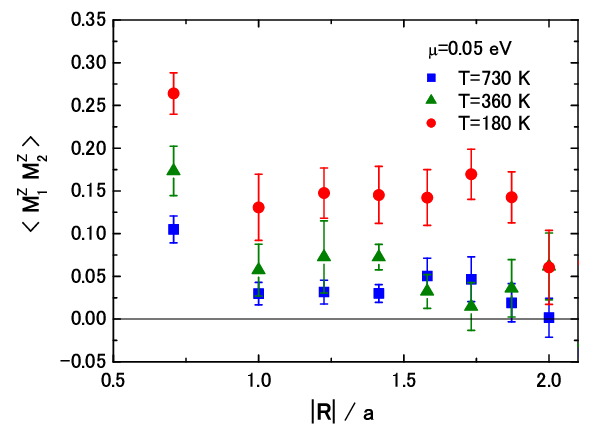}
\caption{(Color online) Inter-impurity magnetic correlation
function $\langle M_1^z M_2^z\rangle$ vs. the impurity separation
$|{\bf R}|/a$ at various temperatures for $\mu$ fixed at 0.05 eV.
\label{fig2}}
\end{center}
\end{figure}

A multi-orbital Haldane-Anderson model 
is given as follows \cite{Haldane1976}:
\begin{eqnarray}
{\cal H} &=& \sum_{{\bf k},\alpha,\sigma} 
(\epsilon_{{\bf k}\alpha}-\mu) 
 c^{\sigma \dagger}_{{\bf k}\alpha} 
 c^{\sigma}_{{\bf k}\alpha} 
+ \sum_{i,\xi,\sigma} (E_{d\xi}-\mu) 
 d^{\sigma \dagger}_{i \xi}
 d^{\sigma}_{i \xi} \nonumber \\
&+&\!\sum_{{\bf k},\alpha,i,\xi,\sigma}
 \!(V_{{\bf k}\alpha,i\xi} c^{\sigma \dagger}_{{\bf k} \alpha} 
 d^{\sigma}_{i \xi}\!+\!{\rm h.c.})\!+\!U \sum_{i,\xi} 
 d^{\uparrow \dagger}_{i \xi}
 d^{\uparrow}_{i\xi}
 d^{\downarrow \dagger}_{i \xi} 
 d^{\downarrow}_{i \xi}. \label{eq:HAM} 
\end{eqnarray}
Here, $c^{\sigma \dagger}_{{\bf k}\alpha}$ 
($c^{\sigma}_{{\bf k}\alpha}$) creates (annihilates) a host electron with
wavevector ${\bf k}$ and spin $\sigma$ in the valence or
conduction bands denoted by $\alpha$, and
$d^{\sigma \dagger}_{i \xi}$ ($d^{\sigma}_{i \xi}$) is a creation
(annihilation) operator for a localized electron at impurity
$d$ orbital $\xi$ located at transition-metal site $i$. 
The first term in Eq.~(\ref{eq:HAM}) represents 
a kinetic energy $\epsilon_{{\bf k} \alpha}$ of 
 a host ${\bf k} \alpha$ electron, and 
the second term describes a bare onsite energy $E_{d \xi}$ 
of the impurity and $\mu$ is a chemical potential. 
The third term specifies impurity-host hybridization 
 $V_{{\bf k} \alpha, i \xi}$ and the last term represents 
the onsite repulsion $U$ at the impurity site. 
In the present study, we keep only diagonal terms in the onsite interaction  
by neglecting the off diagonal terms in the interaction 
(i.e., Hund's rule coupling). 

When a Mn atom is substituted in place of a Ga site, the five 3$d$ orbitals 
of the Mn ion are split into the three-fold degenerate $t_{2g}$ orbitals 
and the two-fold degenerate $e_g$ orbitals by the tetrahedral crystal field. 
In the present Mn$^{2+}$ case with a 3$d^5$ configuration, 
the $e_g$ orbitals are fully occupied and 
therefore are inactive for the magnetic properties. 
In contrast, the three-fold $t_{2g}$ orbitals 
being a partially-filled state are active 
for the low-energy magnetic excitations.  

The onsite repulsion was set to $U$=4.0 eV following 
a photoemission spectrum \cite{Okabayashi1998} 
 and this value is very close to theoretical values 4-5 eV
 for metallic Mn \cite{Ref_Ferdi}.  
We note that both $U$ and $E_{d\xi}$ can shift the energy of the impurity bound state.
Experimentally, it is widely agreed that an impurity band exists in the dilute limit of (Ga,Mn)As at 110 meV above the valence band \cite{Jungwirth2007}.
To realize this situation, we found that $E_{d\xi}$ should be set to $-$2.0 eV.

Estimations for the hybridization parameters $V_{{\bf k}\alpha,i\xi}$ 
in Eq.~(\ref{eq:HAM}) need careful treatments. 
Empirical estimations with the Slater-Koster table fail to describe 
the dispersion of the conduction bands, thus leading to a serious failure in the hybridization-parameter estimation. 
\begin{figure}[tbp]
\begin{center}
\includegraphics[scale=0.6]{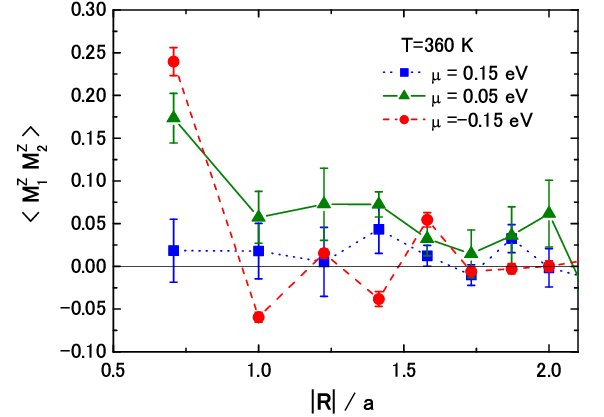}
\caption{(Color online) 
$\langle M_1^z M_2^z\rangle$ vs. $|{\bf R}|/a$ at $T=360$ K for
various values of $\mu$.
\label{fig3}}
\end{center}
\end{figure}
We estimate the hybridization parameter $V_{{\bf k}\alpha,i\xi}$ 
in Eq.~(\ref{eq:HAM}) from {\em ab initio} 
density-functional calculations \cite{Ref_TAPP}. 
The matrix element is given by 
\begin{eqnarray}
V_{{\bf k} \alpha, i \xi} = \langle \psi_{{\bf k}\alpha} 
| {\cal H}_0 | \varphi_{i\xi} \rangle, \label{eq:hy1} 
\end{eqnarray}
where ${\cal H}_0$ is the one-body part of 
Eq.~(\ref{eq:HAM}),
$|\psi_{{\bf k}\alpha}\rangle= c^{\sigma \dagger}_{{\bf k}\alpha}|0\rangle$,
and $|\varphi_{i \xi}\rangle =
d^{\sigma \dagger}_{i \xi}|0\rangle$.
$|\psi_{{\bf k}\alpha} \rangle$ can be expressed with a 
linear combination of Wannier functions as 
\begin{eqnarray}
| \psi_{{\bf k}\alpha} \rangle = \frac{1}{\sqrt{N}} 
\sum_{\mu} \sum_{{\bf R}} a_{\alpha \mu} (\bk) e^{i {\bf k} {\bf R}} 
| w_{{\bf R} \mu} \rangle, \label{eq:Bloch} 
\end{eqnarray}
where $|w_{{\bf R} \mu} \rangle$ is the Wannier orbital 
at the lattice ${\bf R}$ and the label $\mu$ specifies the type of 
the orbitals. 
In the present GaAs case, the $\mu$ specifies 
the four types of $sp^{3}$ orbitals centered at a Ga or As site. 
Expansion coefficients \{$a_{\alpha \mu} (\bk)$\} are obtained 
from solving the equation ${\cal H}_{cry} | \psi_{{\bf k}\alpha} \rangle = \epsilon_{{\bf k}\alpha} | \psi_{{\bf k}\alpha} \rangle$.
By substituting Eq.~(\ref{eq:Bloch}) into Eq.~(\ref{eq:hy1}), 
we obtain the form of 
\begin{eqnarray}
V_{{\bf k} \alpha, i \xi} =  \frac{1}{\sqrt{N}} 
\sum_{\mu} \sum_{{\bf R}} a^{*}_{\alpha \mu} (\bk) 
e^{-i {\bf k} {\bf R}} 
\langle w_{{\bf R} \mu} | {\cal H}_0 | \varphi_{i \xi} \rangle. \label{eq:hy2} 
\end{eqnarray}

For an actual evaluation of the matrix element 
$\langle w_{{\bf R} \mu} | {\cal H}_0 | \varphi_{i \xi} \rangle$, 
we utilize a supercell calculation: 
We first arrange a supercell containing 
$3\!\times\!3\!\times\!3$ fcc primitive cells, 
where a Ga atom located at the origin in the supercell 
is replaced with an Mn atom. 
We then make maximally localized Wannier 
functions \cite{Marzari1997} 
for this system \cite{Ref_supercell}.  
The width of an energy window was set in order to 
include all the valence-bonding and conduction-anti-bonding states. 
The total number of the resulting Wannier functions are 217  
in which there are 212 GaAs Wannier orbitals \{$|w_{n \mu} \rangle$\}
 with $\mu$ and $n$ specifying four types of the $sp^3$ orbitals 
and sites of Ga or As, respectively, and five Mn-3$d$ 
Wannier orbitals \{$|w_{i \xi} \rangle$\} sitting on the origin site $i$. 
Then, we calculate the matrix element 
$\langle w_{n \mu} | 
{\cal H}_{KS}^{3\times3\times3} | 
w_{i \xi} \rangle$ with ${\cal H}_{KS}^{3\times3\times3}$ 
being the Kohn-Sham Hamiltonian for the supercell.
This matrix element is used instead of 
$\langle w_{ {\bf R} \mu} | {\cal H}_0 | \varphi_{i \xi} \rangle$ 
in Eq.~(\ref{eq:hy2}) for ${\bf R}\neq i$. 
We note that the error due to this replacement is scaled by $\frac{1}{M}$
with $M$ being the total number of the atoms in the supercell.  
For ${\bf R}=i$, 
$\langle w_{{\bf R} \mu} | {\cal H}_0 | \varphi_{i \xi} \rangle$ 
is expected to be negligibly small, since wave-function 
symmetry of $|w_{{\bf R} \mu} \rangle$ and
$|\varphi_{i \xi} \rangle$ are different.
We found that the hybridization parameters are rather short-ranged; these are nearly zero except for the nearest and next-nearest sites; for example, the parameters between the Mn-$t_{2g}$ orbitals and 
the nearest As-$sp^3$ orbitals are 0.16-0.74 eV, while the values between the $t_{2g}$ orbitals and the next-nearest Ga-$sp^3$ orbitals are 0.06-0.14 eV. 
Notice that, if the empirical Slater-Koster estimation is employed, the latter next-nearest values are zero.

We perform the quantum Monte Carlo simulation for 
a one- or two-impurity Haldane-Anderson models, 
in which we use the Hirsch-Fye QMC algorithm \cite{Hirsch1986} for
obtaining the electronic correlations in the one- and two-impurity
Haldane-Anderson models. In particular, we study with QMC the
magnetic correlations and the local density of states around the impurity site. We
define the magnetization operator at the $\xi$ orbital of the
$i$'th impurity site as $M^z_{i\xi} = n_{i\xi\uparrow} -
n_{i\xi\downarrow}$, and at site ${\bf r}$ of the host as
$m^z({\bf r}) = n_{\uparrow}({\bf r}) - n_{\downarrow}({\bf r})$.
Here, $n_{i\xi\sigma}$ and $n_{\sigma}({\bf r})$ are the
spin-dependent number operators at the impurity and host sites,
respectively. For the single-impurity case, we present data on the
average value of the square of the local moment at the impurity
site,
%\begin{equation}
$\langle (M^z)^2 \rangle =  \frac{1}{3} \sum_{\xi=1}^3 \langle (
M^z_{i\xi})^2 \rangle,$
%\end{equation}
where $\xi$ sums over the $t_{2g}$ orbitals, while for the
two-impurity case we show data on the magnetic correlation
function defined by
%\begin{equation}
$\langle M_1^z M_2^z \rangle =  \frac{1}{3} \sum_{\xi=1}^3 \langle
M^z_{1\xi} M^z_{2\xi} \rangle$
%\end{equation}
for impurities located at sites ${\bf R}_1$ and ${\bf R}_2$. In
the single-impurity case, we will also present data on the
impurity-host magnetic correlation function,
%\begin{equation}
$\langle M^z m^z({\bf r}) \rangle =  \frac{1}{3} \sum_{\xi=1}^3
\langle M^z_{0\xi} m^z({\bf r}) \rangle,$
%\end{equation}
where ${\bf r}$ is the host site and the impurity is located at
the origin.

Figure \ref{fig1} shows the square of the impurity magnetic moment $\langle
(M^z)^2\rangle$ as a function of the chemical potential $\mu$ at
various temperatures $T$ for the single-impurity Anderson model.
With decreasing $T$, we observe that a step-like discontinuity
develops centered at $\mu\approx 0.1$ eV, which indicates that the
impurity bound state is located at this energy \cite{Bulut2007}. Within the
Hartree-Fock picture \cite{Ichimura2006,Bulut2007}, the
impurity-host hybridization creates both the impurity bound state (IBS) and the host
split-off state at $\omega_{\rm IBS}$. For $\mu < \omega_{\rm IBS}$, the
hybridization with the valence band reduces the impurity magnetic
moment. When $\mu$ is increased to above $\omega_{\rm IBS}$, the
polarization of the split-off state cancels the polarization of
the valence band, which in turn decreases the screening of the
impurity moment. This is the reason for why $\langle
(M^z)^2\rangle$ increases at $\omega_{\rm IBS}$.

\begin{figure}[tbp]
\begin{center}
\includegraphics[scale=0.6]{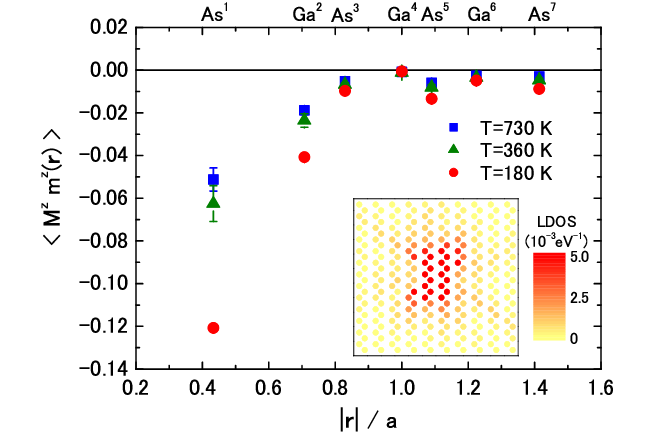}
\caption{(Color online) Impurity-host magnetic correlation
function $\langle M^z m^z({\bf r})\rangle$ vs $|{\bf r}|/a$ for
$\mu=0.05$ eV and various temperatures in the single-impurity
case. (Inset) Local density of states of the host electrons in the
$(011)$ plane. Here, the impurity is located at the next
layer, and the circles indicate the position of the Ga and As
sites. The size of this figure is $5\times 5.6\,$nm$^2$, and these
results are for $T=360$ K and $\mu=0.05$ eV.\label{fig4}}
\end{center}
\end{figure}

Figure \ref{fig2} shows the impurity-impurity magnetic correlation
function $\langle M_1^z M_2^z\rangle$ {\em vs} $|{\bf R}|/a$ at various
$T$ for the two-impurity Anderson model. Here, $|{\bf R}|=|{\bf
R}_1-{\bf R}_2|$ is the impurity reparation and $a$ is the GaAs
lattice constant. In addition, $\mu$ is fixed at $0.05$ eV so that
it is located between the top of the valence band and the impurity bound state. In
Fig. \ref{fig2}, we observe that the ferromagnetic correlations become stronger and
extended in real space, as $T$ decreases down to 180 K, which is
in the physically relevant temperature regime for (Ga,Mn)As \cite{Ohno1992}.
Figure \ref{fig3} shows $\langle M_1^z M_2^z\rangle$ {\em vs} $|{\bf R}|/a$ at
360 K for $\mu=-0.15,\,0.05,\, 0.15$ eV. For $\mu=-0.15$ eV, the
chemical potential is located in the valence band. In this case,
the system is metallic and we observe
Ruderman-Kittel-Kasuya-Yosida (RKKY) type oscillations in the
magnetic correlations. The ferromagnetic coupling disappears for $\mu=0.15$
eV, where the impurity bound state becomes occupied. These results indicate that
the chemical potential should be located between the impurity bound state and the
valence band for obtaining extended ferromagnetic coupling between the
impurity moments. This case corresponds
to the insulating state of (Ga,Mn)As found in the dilute limit
\cite{Jungwirth2007}. In addition, the strong $\mu$,
$R$ and $T$ dependence of $\langle M_1^z M_2^z\rangle$ observed in
Figs.~\ref{fig3} and ~\ref{fig4} makes it difficult to determine {\it
phenomenologically} the magnetic coupling in DMS materials,
instead requiring approaches which treat the band structure and
correlation effect on an equal footing and exactly such as the
DFT+QMC method used here.

In order to determine the origin of the ferromagnetic correlation between
the impurities, we present in Fig.~\ref{fig4} the impurity-host magnetic
correlation function $\langle M^z m^z({\bf r})\rangle$ vs $|{\bf
r}|/a$ at various $T$ for the single-impurity Haldane-Anderson
model. These results are for $\mu=0.05$ eV in which case the
inter-impurity ferromagnetic correlation is strong. We observe that the
impurity-host magnetic coupling is anti-ferromagnetic and it
enhances with decreasing temperature. When the impurity bound state is
occupied ($\mu = 0.15$ eV), we find that the impurity-host anti-ferromagnetic
magnetic coupling becomes much weaker (not shown here) and the
inter-impurity ferromagnetic correlations collapse. Hence, the anti-ferromagnetic
impurity-host coupling generates the indirect ferromagnetic coupling between
the impurities. 

Finally, in the inset of Fig.~\ref{fig4}, we show local density of states around the Mn
site defined by
%\begin{equation}
$N({\bf r}) = -\frac {d}{d\mu} \langle n({\bf r}) -
n(\infty)\rangle\bigg|_{\mu=0.05\,{\rm eV}},$
%\end{equation}
where $n({\bf r})$ is the electron number operator at site ${\bf
r}$. The highly anisotropic distribution of local density of states around Mn is
consistent with the STM measurements
\cite{Yakunin2004,Celebi2008}, which again emphasizes the
necessity of including both the band structure and correlation
effect \cite{Tang2004,Fiete}.

In summary, we have investigated the magnetic properties of the
dilute magnetic semiconductors (Ga,Mn)As in the dilute limit within a combined approach of the density functional theory and the quantum Monte Carlo method. For this purpose, we have mapped the electronic
state of the system to the multi-orbital Haldane-Anderson model by
using the density functional theory with maximally-localized Wannier functions. For
calculating the magnetic properties, we have used the Hirsch-Fye
quantum Monte Carlo method, which does not exhibit a fermion
sign problem. Within this approach, we have correctly reproduced
the experimental value for the energy of the impurity band of
(Ga,Mn)As in the dilute limit.
The numerical results emphasize the
role of the impurity bound state in producing the ferromagnetic correlations. We have found
that, when the chemical potential is between the top of the
valence band and the impurity bound state, the inter-impurity ferromagnetic correlation
originating from the anti-ferromagnetic impurity-host coupling is strong and
extended in real space. This case corresponds
to the insulating state found in the low-doping regime of
(Ga,Mn)As. In addition, we have observed a highly anisotropic
distribution of local density of states around Mn which is consistent with the STM
measurements. The agreements found with the experimental data on
the dilute limit of (Ga,Mn)As suggest that the proposed approach is
a reliable method for investigating and designing other DMS
materials.

We thank H. Ohno and M. E. Flatt\'{e} for valuable discussions.
This work was supported by the NAREGI Nanoscience Project and a
Grand-in Aid for Scientific Research from the Ministry of
Education, Culture, Sports, Science and Technology of Japan.
We thank the Supercomputer Center, Institute for Solid State Physics, University of Tokyo for the use of facilities.


\begin{thebibliography}{99}
\bibitem{Ref_supercell}
 Supercell Wannier functions were constructed with a
 5$\times$5$\times$5 $k$-points sampling.
\end{thebibliography}
\end{document}